\documentclass[sigconf]{acmart}

\copyrightyear{2021}
\acmYear{2021}
\setcopyright{acmcopyright}\acmConference[EASE 2021]{Evaluation and Assessment
in Software Engineering}{June 21--23, 2021}{Trondheim, Norway}
\acmBooktitle{Evaluation and Assessment in Software Engineering (EASE 2021), June
21--23, 2021, Trondheim, Norway}
\acmPrice{15.00}
\acmDOI{10.1145/3463274.3463326}
\acmISBN{978-1-4503-9053-8/21/06}

\usepackage{soul}

\usepackage{tabularx} 

\usepackage{multirow}

\usepackage{float}

\usepackage{booktabs}
\usepackage{tabu}
\usepackage{xcolor}
\usepackage{url}

\begin{document}
\title{Influence of Roles in Decision-Making during OSS Development --- A Study of Python}
\author{Pankajeshwara Nand Sharma}
\email{pankaj.sharma@postgrad.otago.ac.nz}
\affiliation{%
  \institution{University of Otago}
  \streetaddress{P.O. Box 1212}
  \city{Dunedin}
  \state{Otago}
  \country{New Zealand}
  \postcode{43017-6221}
}

\author{Bastin Tony Roy Savarimuthu}
\email{tony.savarimuthu@otago.ac.nz}
\affiliation{%
  \institution{University of Otago}
  \streetaddress{P.O. Box 1212}
  \city{Dunedin}
  \state{Otago}
  \country{New Zealand}
  \postcode{43017-6221}
}

\author{Nigel Stanger}
\email{nigel.stanger@otago.ac.nz}
\affiliation{%
  \institution{University of Otago}
  \streetaddress{P.O. Box 1212}
  \city{Dunedin}
  \state{Otago}
  \country{New Zealand}
  \postcode{43017-6221}
}

\renewcommand{\shortauthors}{Sharma, et al.}

\begin{abstract}

%
%
%
Governance has been highlighted as a key factor in the success of an Open Source Software (OSS) project.   
%
It is generally seen that in a mixed meritocracy and autocracy governance model, the decision-making (DM) responsibility regarding what features are included in the OSS is shared among members from select roles; prominently the project leader. However, less examination has been made  
whether members from these roles are also prominent in DM discussions and how decisions are made, to show they play an integral role in the success of the project. We believe that 
to establish their influence, it is necessary to examine not only discussions of proposals in which the project leader makes the decisions, but also those where others make the decisions. Therefore, in this study, we 
examine the prominence of members performing different roles in: (i) making decisions, (ii) performing certain social roles in DM discussions (e.g., discussion starters), (iii) contributing to the OSS development social network through DM discussions, and (iv) how decisions are made under both scenarios. 
We examine these aspects in the evolution of the well-known Python project. We carried out a data-driven longitudinal study of their email communication spanning 20 years, comprising about 1.5 million emails. These emails contain decisions for 466 Python Enhancement Proposals (PEPs) that document the language's evolution. Our findings make the influence of different roles transparent to future (new) members, other stakeholders, and more broadly, to the OSS research community. 
\end{abstract}





\maketitle

\section{Introduction}\label{sec:introduction}

%


The success of OSS has attracted much attention in both research and managerial practice. These volunteer-based communities are sustained, based on their governance approach to which decision-making processes are crucial \cite{lerner2002some,fleming2007brokerage, giuri2008explaining}. 
While there are good examples of successful OSS projects, e.g., Linux, OpenOffice, and Python, there are many projects which are not successful (\cite{ehls2017open}\cite{lee2019longitudinal}). 

Recently, researchers have investigated the influence of roles, specifically leaders, in Open Source Software Development (OSSD) to understand what contributes to the success or failure of OSS projects. 
Leaders' ``task-oriented leadership behaviors in forms of technical contributions and relation-oriented leadership behavior'' have been identified as major contributors to success \cite{mu2019role}. 
Similarly, leaders with more contributions usually have more expertise and provide satisfactory feedback, thus helping to attract more participants \cite{moon2008role}. They can initiate dialogues, prompt feedback, and shape how members discuss a topic \cite{huffaker2010dimensions}. They also inspire others to ``talk'' \cite{panteli2016leaders} and engage in conversations \cite{johnson2015emergence}. Their findings also suggest that when choosing OSS projects to contribute to, participants should pay particular attention to the project leader.   


In addition, studies of the OSS group decision-making (GDM) processes 
(e.g., \cite{sharma2020mining, mockus2002two,  mockus2000case, mrowka2015decision}) have  showed initial evidence that the sharing of DM responsibility is likely to vary across different OSS communities. 
Recent studies have pointed out that OSS leadership is more ``distributed'' rather than individualistic (\cite{thapa2020evaluating, gronn2002distributed}). Also, leaders can be silent during discussions, while encouraging other members to share their opinions and facilitate information exchange instead of posting many messages \cite{panteli2016leaders}. 
While the success of OSS projects is normally attributed to their leaders, the fact that this leadership is distributed means that 
the positive performance of a project can also be attributed to the core participants who conduct day-to-day guidance, direction, and control of discussions during DM. 

Recently, the influence of OSS project leaders has come under the microscope. Linus Torvalds, the leader of the Linux project, took temporary leave of absence to focus on understanding how to respond to people's emotions \cite{riley2018}. 
Conversely, Python's \textit{Benevolent Dictator For Life (BDFL)} permanently stepped down from his post  over a proposal \cite{steven2018}, and stated that he had never had to fight so hard to get a proposal accepted.

In terms of trying to understand the overall influence of prominent OSS members in different roles in DM discussions and how decisions are made, 
most previous research has only provided snapshots of their involvement (\cite{eseryel2013action, li2012leadership, johnson2015emergence, barcellini2006users, barcellini2005study,mockus2002two}). Thus, there is a need to examine longitudinal data (using a mix of qualitative and quantitative approaches) to establish the influence of these members in various roles who play an integral part in the project's success,  particularly those projects that employ a mixed governance model during their evolution. At a high-level, this paper investigates the question: 
\textit{How are members in different prominent OSSD roles involved in daily DM discussions on OSS enhancements?} 
Particularly, what is the overall influence of members in these roles during DM in terms of: (a) who makes decisions, (b) their social roles during DM discussions (e.g., initiating discussions and answering questions), (c) their prominence in the social network, and (d) considerations of their preferences when decisions are made, including comparison of these four aspects when other members make those decisions.

These research gaps can be addressed by scrutinizing and reflecting on the DM practices of successful OSS projects. 
This work bridges the above-mentioned gaps by discovering the influence of decision-makers during DM discussions  through longitudinal 
data-driven analysis of the successful Python OSS project, which has been noted to have a good governance model \cite{wang2015comparative}, and has been studied by several researchers (\cite{barcellini2005thematic,  barcellini2005study, sack2006methodological, mahendran2002serpents, barcellini2006users, keertipati2016exploring,savarimuthu2016process}). Discussions about 466 Python Enhancement Proposals from 1.5 million emails are studied to reveal the nature of the roles in DM.

This paper is organized as follows: Section~\ref{sec:backgroundRQs} provides the background on relevant work and presents the research questions,  Section~\ref{sec:methodology} presents the methodology used to extract the influence of Python members, Section~\ref{sec:results} presents the results while Section~\ref{sec:discussion} discusses the contributions, and Section~\ref{sec:conclusion} presents our conclusions.

\section{Background and Research Questions} \label{sec:backgroundRQs}

Three factors that have a detrimental effect on the members of an OSS community are a change of leadership and governance structures \cite{ehls2017open}, a lack of DM process transparency, and consent in the DM process \cite{scacchi2008governance}. A cross-cutting concern in these three factors is the nature of roles performed by individuals during DM and the extent to which these roles positively or negatively influence the governance of a project, including the key DM steps.



OSS governance is often depicted as an \emph{onion model} \cite{nakakoji2002evolution} that shows the hierarchy of roles, with the significant roles identified in the middle of the onion (core, with few members such as project leaders and core-developers) and the outer layers represent non-core, yet useful activities performed by many members (e.g. bug reporters and software users). 

A limitation of the current literature in applying the onion model 
to OSS projects with a mixed governance approach  
is that it has only had a qualitative focus, where high level DM models are presented based on interviews and surveys  with stakeholders.  While qualitative approaches may offer valuable insights, they also introduce a bias of exclusively relying on individual opinions \cite{thapa2020evaluating}. Even when the ``actual'' DM models have been studied using a data-driven approach, insights have been fragmented and limited \cite{thapa2020evaluating}. 


This influence of members can only be established by examining their day-to-day activities during DM
--- an approach belonging to the practice theory perspective, which emphasizes the actual practices employed in everyday community processes \cite{orlikowski2002knowing}. 
We thus examine the following four aspects using a data-driven study of Python email archives.

\emph{(a) Decision-makers:} \label{sec:whomakesdecisions}
We first focussed on establishing members who have decision-making \textit{roles} regarding OSS enhancement proposals. 
There are two challenges.
First, members progress in their roles within the community \cite{ye2003toward}, 
and thus the actual DM contributions of each member in each role has to be established. 
Second, a participant can perform different roles for different proposals (e.g., author in one but only a participant in another). 
This needs to be accommodated in the analysis. 
Thus, our first research question is:
\textbf{RQ1: Members in which Python roles make PEP decisions?}

\vspace{.15cm}
\emph{(b) Decision-makers' social roles in DM discussions:} 
Next, we associate decision-makers with their 
\textit{social roles} within OSS enhancement DM discussions. Social roles are useful because researchers have found the OSS design process is influenced by the connections members have within their community \cite{barcellini2005study, barcellini2005thematic}. 
People in these roles 
initiate most discussions (such as question people), reply to a lot of top-level posts (such as answer people), and (iii) have messages attracting a high number of replies, and are ever present in discussions. 
These social roles are examined in our next research question:
\textbf{RQ2: Do members in certain Python roles participate more than others in DM discussions in terms of: (i) total posts, 
(ii) seeding posts, 
(iii) replying to posts, (iv) receiving replies, and (v) involvement in threads, 
and does this participation change when someone other than the BDFL makes the decisions?}

\vspace{.15cm}
\emph{(c) Decision-makers' influence in OSS social network during DM discussions:}
Researchers have used social network anaylsis (SNA) to highlight the influence and positions of key members, describing the role participants play in influencing decisions \cite{wonodi2012using}. 
Current works have identified core developers and corresponded them with decision-makers (\cite{sureka2011using,long2006understanding,crowston2006core}), 
to point out that rather than a member's expertise, his/her position in the OSS community structure is more likely to allow for innovation \cite{dahlander2012core}.
Long and Siau \cite{long2007social} concluded that the social structure of the OSS project affects DM in communities, while in our previous work (\cite{sharma2017boundary}), we found that boundary spanners corresponded with the actual decision-makers. Meneely at al. \cite{meneely2010improving} 
concluded that the central developers in a developer network have a high likelihood of being approvers. 

None of this prior work, however, has attempted to identify how much influence the OSS project leader (in a mixed governance model) has in the OSSD network during DM discussions and how this changes when others make the decisions. We use the theory of ``networked influence'' \cite{lee2019longitudinal} to investigate network patterns where the leader influences the behaviour of others to sustain a successful OSSD network. 
Thus, our next research question is:
\textbf{RQ3: What is the influence of members in various Python roles in the decision-making social network and does it change when someone other than the BDFL makes the decisions?}


\vspace{.15cm}
\emph{(d) How decisions are made:} \label{sec:howdecisonsaremade}
Prior works have employed qualitative approaches such as interviews to understand OSS project governance (\cite{markus2007governance,jensen2010governance, mrowka2015decision, Hanganu2012voting}). 
It is only recently that researchers have started to engage data-driven quantitative approaches.  
Various works (\cite{sharma2020mining, savarimuthu2016process,keertipati2016exploring,sharma2017investigating, sharma2021rationale} have unearthed the ``actual'' DM processes as opposed to the ``advertised'' ones in projects with a mixed governance approach (i.e., consensus-based with dictatorship) with varied levels of detail. 
However, the focus of most of these works has been on decisions themselves (e.g., \cite{li2020automatic}), and not on ``how'' they are made \cite{razavian2019empirical, bhat2020evolution}. Thus, the influence of OSS participants on how design decisions are made remains underexplored, 
which leads to our final research question:
\textbf{RQ 4: How do members from various Python roles influence PEP decision-making 
and does it change when someone other than the BDFL makes the decisions?}




\section{Methodology} 	\label{sec:methodology}

To answer these research questions, 
we employed a range of methods including comparative analysis, social network analysis, and content analysis of discussions for all proposals in the Python OSS project. 
Python was chosen because it operates based on a mixed governance model including elements of both meritocracy and autocracy \cite{sharma2020mining}. It is also widely used and popular \cite{popularity2020}, has a reputation for following good governance practices \cite{wang2015comparative}, has garnered significant research interest \cite{keertipati2016exploring,savarimuthu2016process}, and makes data available to the public (e.g., email archives and commit repositories \cite{dam2015mining, keertipati2016exploring}). 

The Python language is modified and evolves by means of formal \emph{Python Enhancement Proposals (PEPs)}. These PEP documents move through various states at different stages of the DM process. Once the \textit{draft} is ready, it is circulated within the Python community (developers and users) for discussion. The PEP will then eventually be \textit{accepted} or \textit{rejected}. This process 
is outlined in PEP  1\footnote{https://www.python.org/dev/peps/pep-0001/}. Our methodological steps are described next. 


 \textbf{Step 1: Data Extraction.} The first step was to download all 466 PEPs (including their metadata) and all email messages related to them. 
 We obtained the PEPs from Python's Github repository
 and stored them in a 
 database. 
  We considered data from March 1995 to 12th July 2018 --- the date when the BDFL resigned, marking the end of the benevolent dictatorship governance model. We then extracted all individual email messages from the Python email archives, and stored them in a MySQL database. We used six mailing list archives (\emph{python-dev}, \emph{python-ideas}, \emph{python-commits}, \emph{python-checkins}, \emph{python-list}, and \emph{python-patches}) as they are the leading forums for Python developer discussions, with \emph{python-dev} being acknowledged as the primary forum. The resulting dataset contained a total of 1,553,564 email messages. During this step, we also assigned PEP numbers to email messages where 
 PEP numbers or titles were mentioned within the message, or were replies to these messages. 

\textbf{Step 2: Selecting PEP-related messages.} From all the email messages stored in the database, we selected only those that were assigned a PEP number. 
This left a total of 91,311 unique email messages related to these 466 PEPs which we analysed in our dataset. 

\textbf{Step 3: Identity matching.} 
To ascertain all email messages from the same author, we used identity matching techniques \cite{wiese2016mailing}. 
We first applied the approach proposed by Bird et al. \cite{bird2006mining}, which used several combinations of similarity measures for identity matching. We then applied a cumulative identity similarity approach proposed in \cite{bird2008latent} to the output from the first approach. 
To find messages written by the PEP author or BDFL Delegates, we used the PEP metadata available online. For an example, see the metadata for PEP 1 in the link mentioned in footnote 1.


\textbf{Step 4: Identifying roles (to answer RQ1).}
To investigate members and their Python roles associated with DM for each PEP in each PEP type,  
we first established the dates when they were inducted into two specific roles (\emph{PEP editors} using  \cite{python2018peppurpose}\footnote{This is the last version of PEP~1 before the BDFL stepped down and a steering council was elected.} and \emph{Core Developers} using the public list of core developers \cite{pythonofficialcoredevlistpointer,pythondevguidelines}. 
Then, we identified another two roles: \textit{PEP authors} and \textit{PEP decision-maker} (when an individual in any of the roles had been made, or volunteered as, the BDFL Delegate) based on the metadata available about the PEP.
The latter information is available through PEP metadata and in email messages. Finally, a \emph {PEP contributor} is one who has contributed to discussions, but does not belong to any of the other roles. Also, these contributors include members who are not authors or delegates for a particular PEP. 
\textbf{Step 5: Examining social roles (RQ2).}  To examine each member's involvement in DM discussions (social roles) for PEPs, we focused on five aspects as mentioned in RQ2.


\textbf{Step 6: Studying prominence of roles in social networks (RQ3).}  The influence of members can be shown using a social network. Using ``To'' and ``From'' fields, we identified the communications of the BDFL, PEP author, and BDFL Delegate roles. These were then imported into the NodeXL software to construct a social network of Python members' interaction and to compute the two following SNA metrics. 

\textit{Betweenness centrality} measures the centrality of a node in the network by considering the number of the shortest paths the node is part of. It demonstrates high levels of social status, power, and managerial roles in both open source and commercial contexts (\cite{bird2008latent} and [18]).  \textit{Eigenvector centrality} measures how well a node is connected to other well connected nodes. Thus, it is a fitting measure of a member's influence in Python.  

\textbf{Step 7: Studying the influence of roles in DM (RQ4).}  We conducted an in-depth manual analysis of email messages (spanning multiple mailing lists) relating to only the 248 PEPs which were either \textit{accepted} or \textit{rejected}. The first author read the corresponding messages and identified the rationale behind how each PEP was decided using a custom-built GUI tool\footnote{\label{fn:repository}
A screenshot of the tool used to analyze messages and other supplementary data can be viewed in the repository \url{https://github.com/sharmapn/influenceOfRoles}}.
Of these 248 PEPs, we curated 300 rationale sentences relating to 193 PEPs that had an explicitly stated rationale behind their decisions. 
All rationale sentences in the dataset were verified by the second author (i.e., 100\% consensus).	

\section{Results}	\label{sec:results}


\subsection{RQ 1: Python roles responsible for DM} 
\label{sec:rolesonpeptypes}

\par 

We identified four distinct roles in Python development: 
(i) \textit{contributors} who help by reporting bugs, proposing ideas, and improving  documentation; (ii) \textit{core developers}, the key task performers who are heavily involved in discussing and shaping new proposals and implementing them (e.g., developing new Python libraries); 
(iii) \textit{PEP editors} who are the overseers of PEP quality; and (iv) \textit{BDFL}, the final decision maker. We were interested in how each of these roles were involved in two main DM related roles (the PEP authorship role, and the Decision-Maker role for a particular PEP). 

\begin{figure}[ht!] 
	\begin{center}
		\includegraphics[width=87mm]{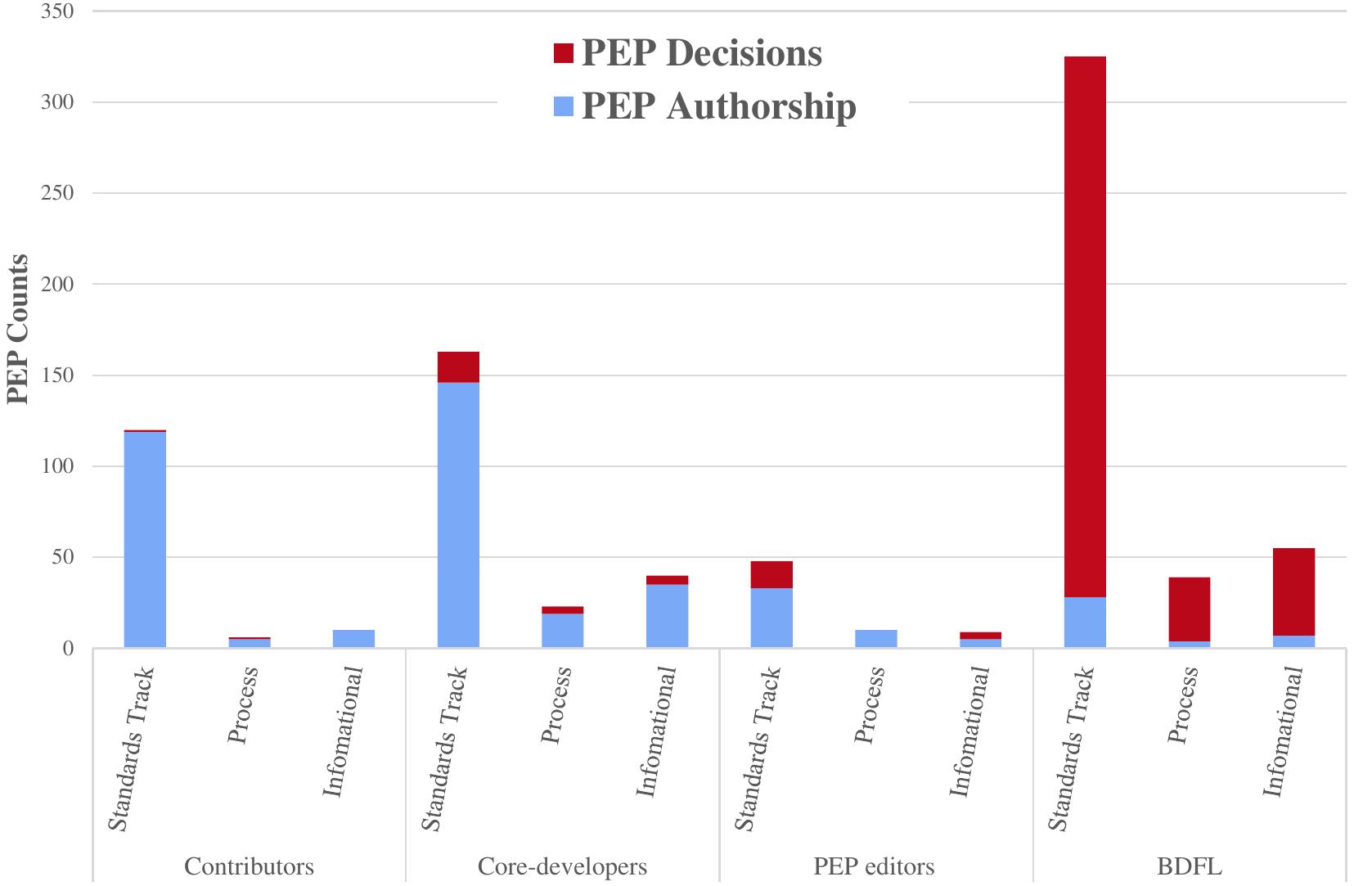}	
	\end{center}
	\caption{PEP authorship and decisions by members in different roles for the three PEP types 
	} 
	\label{fig:pythonrolescontributionbypeptype}
\end{figure}

Fig.~\ref{fig:pythonrolescontributionbypeptype} shows the contributions of different roles to PEP authorship and PEP decisions, broken down by three PEP types (Process, Standards Track and Informational). \textit{PEP authors} who volunteer to champion a new idea by writing a PEP are either contributors or core developers (shown in blue). The number of contributors who are PEP authors are similar to those of core developers. The \textit{PEP decisions} (shown in red), are made by the BDFL or BDFL-delegates (nominees of the BDFL who are members from the first three roles). We found that this responsibility lies mostly with the BDFL, supported to a lesser extent by the senior community members, i.e., the core developers and the PEP editors. 

Contributors do not normally make the final decisions on PEPs as they are relatively new members; indeed, this has only happened twice: PEP 541 \cite{python2017pep541} when it was decided that the BDFL Delegate (a contributor) would only recommend the PEP's acceptance and PEP 484 \cite{python2014pep484}, which was decided by a senior contributor who had only recently accepted the BDFL-Delegate position \cite{python2018e}. Core developers, on the other hand, are active as BDFL delegates across all three PEP types. PEP editors have also served as BDFL delegates who make decisions in Standards Track and Informational PEPs. More Standards Track PEPs have been decided by BDFL delegates than the other two PEP types because the BDFL considered the delegates possessed the most knowledge in certain areas since these PEPs directly concern modifications to the language compared to the other two PEP types. 

There were 51 PEPs decided by someone other than the BDFL. Of these, 35 were Standards Track, 10 were Informational, and 6 were Process PEPs. Of these, almost half (25 PEPs) had the most discussions in the \textit{disutils-sig} mailing list.
From 2014 \cite{lwn2014}, Python packaging-related PEPs were discussed here to fast-track decisions and Nick Coghlan (PEP editor from 2012) decided most of these.

\subsection{RQ 2: Decision-makers' social roles in DM discussions} \label{sec:resultsInfluencePEPDMDiscussions}

Table \ref{tab:Topcontributingmembers} shows the top three Python roles (authors, BDFL and BDFL delegates) and the eight members who are prominent in Python DM discussions
for the 415 PEPs decided by the BDFL. All eight members are Python core developers  \cite{pythondevguidelines,sharma2017boundary}. 
For comparison, we also have included (in italics) the contributions of BDFL, PEP authors and the top two members (Brett Cannon and Nick Coghlan)
in the 51 PEPs where the BDFL Delegates made the decisions.
Note the eight members identified by name have also been found to be prominent boundary spanners (who contribute across different mailing lists) from another work \cite{sharma2017boundary}.
Full email contribution data for all 466 PEPs decided either by the BDFL or a BDFL Delegate and further separation of these contributions by PEP type are available online\textsuperscript{\ref{fn:repository}}. 
Table \ref{tab:Topcontributingmembers} shows the results for questions 2(i-v) in columns 3-7. 

\emph{PEP involvement} --- The BDFL contributes to the most number of PEPs (406) when he makes the decision.  Of the nine PEPs in which the BDFL did not participate: four were eventually \textit{withdrawn},
one PEP is still \textit{active}, 
and one PEP is reserved for future use. 
The PEP authors (as a collective) had the second most involvement, and this was expected as these authors are naturally involved in discussions of PEPs they author. Sometimes PEP authors disappear after proposing a PEP idea which explains the 23 PEPs in which they were not involved in discussions. Two other members contributed to most PEPs --- Brett Cannon and Nick Coghlan.

\emph{Total message posts --- } PEP authors collectively posted the most number of messages. Their contributions across all PEP types are two and a half times more than the next highest contributor, the BDFL. 
The BDFL's strong involvement in DM discussion contributions occurs at different stages of discussions, including messages that summarise what has taken place  so far
or to raise his issues with the PEP. His contributions suggest that he is also driving discussions together with the PEP authors (for each PEP). A third member is Nick Coghlan, a core developer who has been a BDFL Delegate the most times \cite{sharma2017boundary} and who actively contributed to discussions. 

\begin{table}[hb!]
	\centering
	\caption{Top contributing members and roles
        during DM discussions 
		across the 415 and 51 PEPs decided by the BDFL and BDFL Delegates, respectively. 
	}
	\label{tab:Topcontributingmembers}
	\resizebox{\columnwidth}{!}{%
		\begin{tabular}{lrrrrrr}
			\toprule	
			\begin{tabular}{@{}l@{}}\textbf{Member}\\\textbf{or Role}\end{tabular}
			& \begin{tabular}[c]{@{}r@{}} \textbf{PEP}\\ \textbf{Invol.} \end{tabular}
			& \begin{tabular}[c]{@{}r@{}} \textbf{Total}\\ \textbf{Posts} \end{tabular}
			& \begin{tabular}[c]{@{}r@{}} \textbf{Seeding} \\ \textbf{Posts} \end{tabular} 
			& \begin{tabular}[c]{@{}r@{}} \textbf{Replies}\\ \textbf{Sent} \end{tabular} 
			& \begin{tabular}[c]{@{}r@{}} \textbf{Replies} \\ \textbf{Received} \end{tabular}             
			& \begin{tabular}[c]{@{}r@{}} \textbf{Thread}\\ \textbf{Invol.} \end{tabular} 
			\\
            \midrule
			\multirow{2}{*}{PEP Authors} & 392 & 19173 & 8214 & 10959 & 9921 & 9354 \\
			& \textit{51} & \textit{4190} & \textit{1089} & \textit{3101} & \textit{2798} & \textit{3452} \\
            \midrule
			\multirow{2}{*}{BDFL} & 406 & 7757 & 2608 & 5149 & 4608 & 3308 \\
			& \textit{48} & \textit{791} & \textit{248} & \textit{543} & \textit{584} & \textit{328}\\
            \midrule
			\multirow{2}{*}{Brett Cannon} & 338 & 3406 & 1841 & 1565 & 1329 & 2231 \\
			& \textit{34} & \textit{286} & \textit{80} & \textit{206} & \textit{185} & \textit{140}\\
            \midrule
			\multirow{2}{*}{Nick Coghlan} & 302 & 5440 & 1061 & 4379 & 3571 & 1677 \\
			& \textit{51} & \textit{2092} & \textit{428} & \textit{1664} & \textit{1433} & \textit{628} \\
            \midrule
            Barry Warsaw & 274 & 4023 & 2721 & 1302 & 1463 & 2913 \\
            Georg Brandl & 270 & 2539 & 1949 & 590 & 581 & 2083 \\
            Terry Reedy & 209 & 1201 & 397 & 804 & 655 & 596 \\
            Antoine Pitrou & 196 & 1657 & 1083 & 574 & 1300 & 1142 \\
            Greg Ewing & 182 & 1606 & 119 & 1487 & 1026 & 377 \\
            Paul Moore & 174 & 1225 & 205 & 1020 & 996 & 344 \\
		    \midrule
			BDFL Delegate & \textit{51} & \textit{1328} & \textit{283} & \textit{1045} & \textit{1045} & \textit{366} \\
            \bottomrule
			\multicolumn{7}{l}{Italicised values indicate involvement in PEPs decided by BDFL Delegates.} \\ 
	\end{tabular}}
\end{table}


\begin{figure*}[ht!]
	\begin{center}
		\includegraphics[width=180mm]{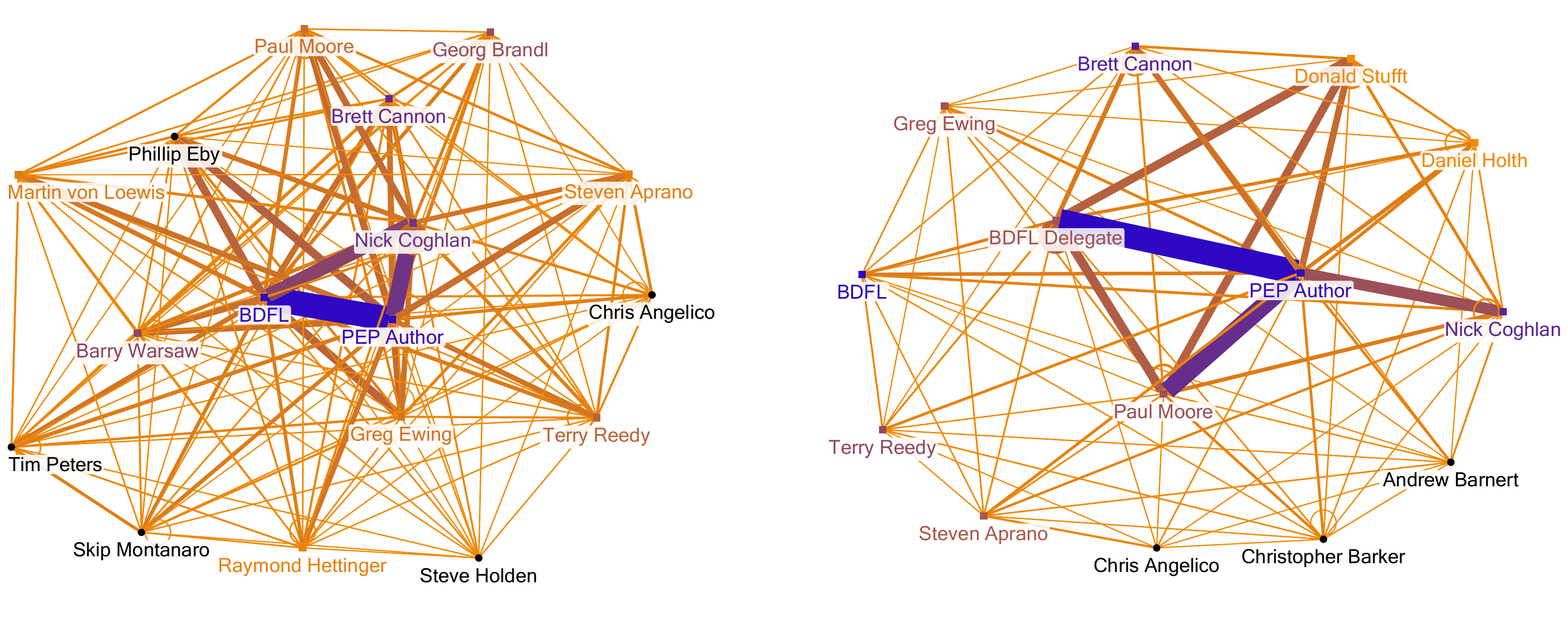}	
	\end{center}
	\caption{A social network showing the influence of the BDFL in the discussions of the 415 PEPs decided by him (left) against the 51 PEPs decided by the BDFL Delegates (right). 
		The vertex colour and size correspond to the number of times a member has been involved in a PEP. The edge width corresponds to the number of connections between two members or roles.
	}
	\label{fig:snabdfl2}
\end{figure*}

\emph{Seeding posts --- } PEP authors collectively are again most prominent. The PEP author (as the PEP champion) is responsible for building consensus within the community, and is generally expected to spearhead PEP discussions. 
The BDFL has the second highest numbers, suggesting he is also spearheading discussions. Barry Warsaw, another core developer, has seeded an almost similar number to that of the BDFL.

\emph{Replies sent --- } PEP authors again are most prominent, which may be due to answering questions from the community. 
Their collective contributions are twice those of the BDFL. The BDFL also shares his knowledge during DM discussions by responding to questions about PEP. His posts also raise issues on the discussion topics present in the seeding posts.

\emph{Replies received --- } These are responses received by members on their message posts.  Again, PEP authors receive the most responses followed by the BDFL. The fact that the BDFL is second implies that he raises important issues which are worth responding to, or that members are interested in obtaining his views, by revealing their positions about issues in their replies. 

\emph{Thread involvement --- } This metric highlights how many distinct discussion threads a member was involved in. Again, we found that PEP authors were the most prominent, followed by the BDFL. 
The BDFL's high thread involvement only highlights his knowledge sharing trait, which is important for sustaining the community.

\begin{table*}[htb!]
	\centering
	\renewcommand\thetable{3}
	\caption{Roles and sample sentences that depict the rationale behind their decisions for \textit{accepted} and \textit{rejected} PEPs. The description of each rationale and their proportions grouped by PEP types are available online.\textsuperscript{\ref{fn:repository}} }
	\label{tab:reasonssentencetable}
	\resizebox{\textwidth}{!}{\begin{tabular}{llllr}
		\toprule
		
		\textbf{Role} & \textbf{State} & \textbf{Rationale} & \textbf{Rationale sentence} & \textbf{PEP}  \\ 
		\midrule
		BDFL & \textit{Acc.} & Consensus & ``The user community unanimously rejected this so I won't pursue this idea any further'' \cite{python2001c} & 259 \\
		PEP Sum. & \textit{Rej.} & No Consensus & ``Although a number of people favored the proposal there were also some objections
		'' \cite{python2007b} & 3128 \\
		Core dev. & \textit{Acc.}& Lazy Consensus & ``If anyone has objections to Michael Hudson's PEP 264: raise them now.'' \cite{python2001b} & 264 \\
		Core dev. & \textit{Rej.} & Rough Consensus & ``Several people agreed, and no one disagreed, so the PEP is now rejcted. [\emph{sic}]'' \cite{python2005b} & 265 \\
		PEP Sum. & \textit{Rej.} & Little Support & ``It has failed to generate sufficient community support in the six years since its proposal.'' \cite{python2001a} & 268 \\
		Core dev. & \textit{Acc.} & Majority & ``Comments from the Community:  The response has been mostly favorable''. \cite{python2002c} & 279 \\
		BDFL & \textit{Rej.}  & No Majority & ``Accordingly, the PEP was rejected due to the lack of an overwhelming majority for change.''\cite{python2003a} & 308 \\
		PEP Sum. & \textit{Rej.} & Inept PEPs & ``This PEP is withdrawn by the author.''\cite{python2002z} & 296 \\
		BDFL & \textit{Acc.} & BDFL Decree & ``After a long discussion, I've decided to add a shortcut conditional expression to Python 2.5.''  \cite{python2005a} & 308 \\
		BDFL & \textit{Acc.} & BDFL PNC	& ``There's no clear preference either way here so I'll break the tie by pronouncing false and true.'' \cite{python2002a} & 285 \\		
		BDFL & \textit{Rej.} & BDFL PM & ``Python is not a democracy.'' \cite{python2004a} & 326 \\
		\bottomrule
		\multicolumn{5}{l}{PEP Sum. = PEP Summary, BDFL PNC = BDFL Pronouncement after No Consensus, BDFL PM = BDFL Pronouncement over Majority}
	\end{tabular}
	}
\end{table*}

\subsection{RQ 3: Influence in the DM Social Network} \label{sec:resultsInfluencePEPDMSNA}
Fig.~\ref{fig:snabdfl2} highlights the prominence of members and the two roles 
in the social network during PEP decision-making when the BDFL made decisions (left) and when the BDFL Delegates made decisions (right). Members are filtered by those having the highest betweenness  centrality. 
Most of the members in both diagrams are Python core developers who have been BDFL Delegates for PEPs 
as previously reported in \cite{sharma2017boundary}. 
The diagram on the right shows the BDFL Delegates both as a collective node and individual nodes. 

We can see that collectively, the PEP authors have the most prominent role in terms of centrality in both social networks.
In discussions about PEPs decided by the BDFL, the BDFL is second most prominent and there is a strong connection (messages exchanged) between the BDFL and PEP authors. These two roles are also central and connected to many core developers in the social network during DM. Nick Coghlan is third most prominent, as he is highly connected to the PEP authors and the BDFL.

In discussions about PEPs decided by BDFL Delegates, BDFL Delegates are highly connected to PEP authors. 
In addition, Nick Coghlan and Paul Moore are more central and strongly connected to the PEP author and the BDFL Delegates. 
The BDFL is not as central in the network, and is not greatly connected to anyone; however, he is still connected with the PEP Authors and BDFL Delegates.

Table 2 shows the SNA metrics of members when PEP decisions were made by the BDFL and the Delegates, respectively. The detailed results of SNA metrics in both cases are contained in the `SNA metrics' sheet within the online repository\textsuperscript{\ref{fn:repository}}. 
Focussing on the two SNA metrics, we found that 
in both cases, the PEP authors had highest values for betweenness and eigenvector centrality. 

\begin{table}[H]
	\centering
	\renewcommand\thetable{2}
	\caption{SNA metrics in PEPs decided by the BDFL and BDFL Delegates}
\label{tab:snametrics}
	\begin{tabular}{|l|r|r|r|r|}
		\hline
		\multirow{2}{*}{\textbf{Role}} & \multicolumn{2}{c|}{\textbf{BDFL}} & \multicolumn{2}{c|}{\textbf{BDFL Delegate}} \\ \cline{2-5}
		& \textit{Betweenness} & \textit{Eigen.} & \textit{Betweenness} & \textit{Eigen.} \\ \hline
		PEP Author   & 737391.753  & 0.012  & 46859.855   & 0.028  \\ \hline
		BDFL         & 261465.886  & 0.010  & \color{blue}4843.832    & \color{blue}0.012 \\ \hline
		Nick Coghlan & 246234.593  & 0.010  & 14696.106   & 0.020  \\ \hline
		BDFL Delegate & N/A  & N/A  & \color{blue}25172.653   & \color{blue}0.022  \\ \hline
	\end{tabular}
\end{table}

The BDFL had the second highest values in the first case, but in the second case, the Delegates and three members had higher values than the BDFL. Of these three members, Nick Coghlan has been the BDFL Delegate the most, while Paul Moore has been on two occasions \cite{sharma2017boundary}. 
Thus, their prominence in these SNA metrics could be attributed to them having high involvement when making decisions. However, the third member, Steven Aprano, had high betweenness centrality, meaning he was neither higher than the BDFL on eigenvector centrality, nor was he high in contributions in the second case (refer to 'Overall Contributions' sheet in the repository\textsuperscript{\ref{fn:repository}}). 

Thus, based on the SNA metrics, apart from the PEP authors, the members who are tasked to make PEP decisions play a more central and influential role in the social network during DM discussions than other members. 


\subsection{RQ 4: Overall Python GDM structure}  \label{sec:pythoncommunitydmstructure}
To find out how Group Decision-Making (GDM) is carried out on PEPs, we examined PEP messages for explanations on the rationale behind their decisions (Section \ref{sec:methodology}). 
In terms of PEP decision-making, a \textit{rationale} refers to the GDM basis on which the final decision on the PEP was made and whether the decision was made using consensus or lazy consensus etc.  In this regard, we found 300 sentences containing rationale behind decisions for 248 \textit{accepted} and \textit{rejected} PEPs.

Table 3 shows examples of rationale sentences corresponding to each of the rationale we manually extracted from the 300 ground truth rationale sentences. A spreadsheet containing the raw data for these sentences and descriptions of the rationale is available online\textsuperscript{\ref{fn:repository}}.
Manually extracting rationale from the Python communication datasets was laborious. We spent three months full-time on the exploratory analysis as it required reading most messages relating to the 245 PEPs that were analysed, with some messages requiring multiple readings to fully understand the nuances.


\begin{figure*}[ht!] 
	\begin{center}
		\includegraphics[width=155mm]{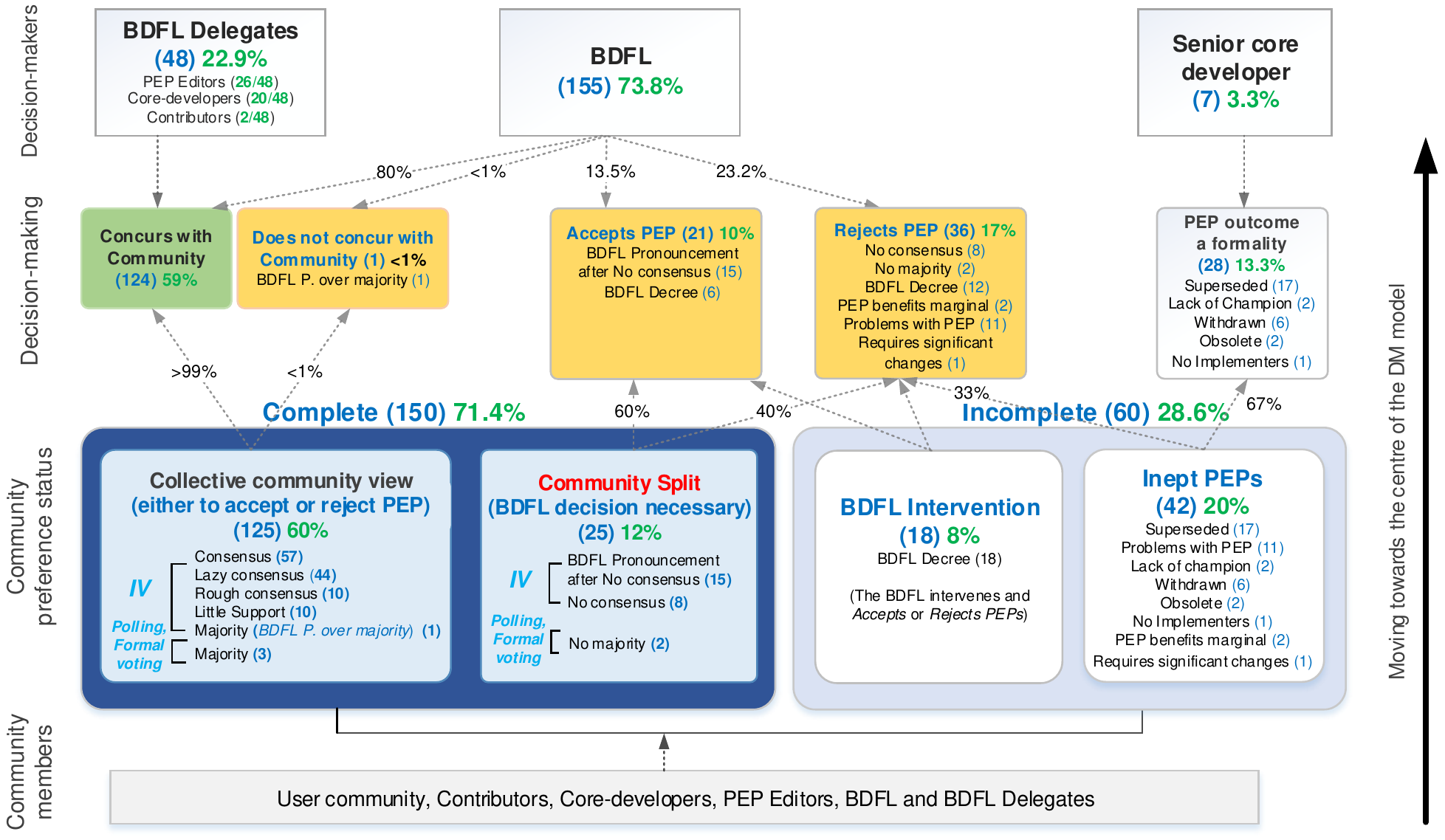}		
	\end{center}
	\vspace{1pt}
	\caption[Overall layered DM structure of the Python OSSD community]{Overall layered PEP DM structure (based on decisions on 248 \textit{accepted} and \textit{rejected} PEPs). 
		Arrows depict the how decisions are made by each entity, including their percentages. Values in green show the percentage of PEP decisions made using that criterion. Dark yellow rectangles represent the BDFL's independent decisions. \textit{IV} refers to Informal Voting.  
	}	
	\label{fig:PythonDMConcept}
\end{figure*}


To evaluate the correctness of our results (i.e., that the identified rationale is correct), we presented our findings to a prominent Python core developer (name withheld) who had been the lead author or co-author of 22 PEPs, had been BDFL Delegate 14 times \cite{sharma2017boundary}, and was a former Python Steering Council member. He replied ``I think your list of reasons looks good''. This demonstrates that the extracted rationale is accurate. 

Based on the data on the rationale (and their proportions) behind the PEP decisions, Fig.~\ref{fig:PythonDMConcept} depicts the overall layered Python DM hierarchy correlating the different sets of participants, community preference gathering outcomes, how decisions are made (emphasising the influence of roles), and the various decision-makers.  
In the figure, moving  the top of the PEP DM structure is the same as moving towards the centre of the onion model. 

For PEPs, community preference status is gathered from community members.
	If the community has \textbf{completed} the preference gathering process, two situations can arise.
		First, a collective community view on a PEP has been established (using \textit{consensus, no consensus, lazy consensus, rough consensus, little support, majority} or \textit{no majority}). In these cases, the BDFL (or his Delegates) 	
			\textbf{\textit{concurred}} in 99\% of the PEPs, which corresponds to 60\% of the PEPs in the dataset, and   
			\textbf{\textit{did not concur}} in less than 1\% of the PEPs. 
		Second, when an overall community view could not be established after discussions (community is split), resulting in either \textit{no consensus} or \textit{no majority},		
		the BDFL made a choice and \textit{accepted} 15 PEPs and \textit{rejected} 10 PEPs. 
		
	If, however, the community preference gathering process remains \textbf{incomplete}, it can be for two reasons.	
		First, during some PEP discussions the BDFL intervened and  accepted six PEPs and rejected 12 PEPs using the BDFL Decree. 
		Second, when it is apparent to the community that some PEPs will not succeed (\textit{Inept PEPs}), the discussions decline and decisions on them are just a formality.  
			In 18 such PEPs, the \textit{\textbf{BDFL intervened}} during PEP discussions and rejected the PEP (mostly disliking the syntax). 
			Sometimes the discussions on a PEP are stalled for some reason and the \textit{\textbf{decision is just a formality (13.3\%)}}. The BDFL, his delegate, or even the trusted senior core developer in charge of maintaining the PEP repository decides on these PEPs. 
While the \textbf{BDFL} makes most of the decisions (73\%) about which PEPs can be accepted into the language, members from other roles can also make these decisions as BDFL Delegates, as highlighted in Section \ref{sec:rolesonpeptypes}. 
	We found three sets of roles representing \textbf{BDFL Delegates}: contributors (4\%),  core developers (55\%), and PEP editors (40\%). We found that these members always sided with the collective view of the community.
	As previously mentioned, some of the inept PEPs 
	lie inactive for a long period without closure, so some \textbf{senior core developers} reject these ``long standing'' PEPs as part of batch cleanup work (3.3\%).  

\begin{figure*}[htb!]
	\begin{center}
		\includegraphics[width=140mm]{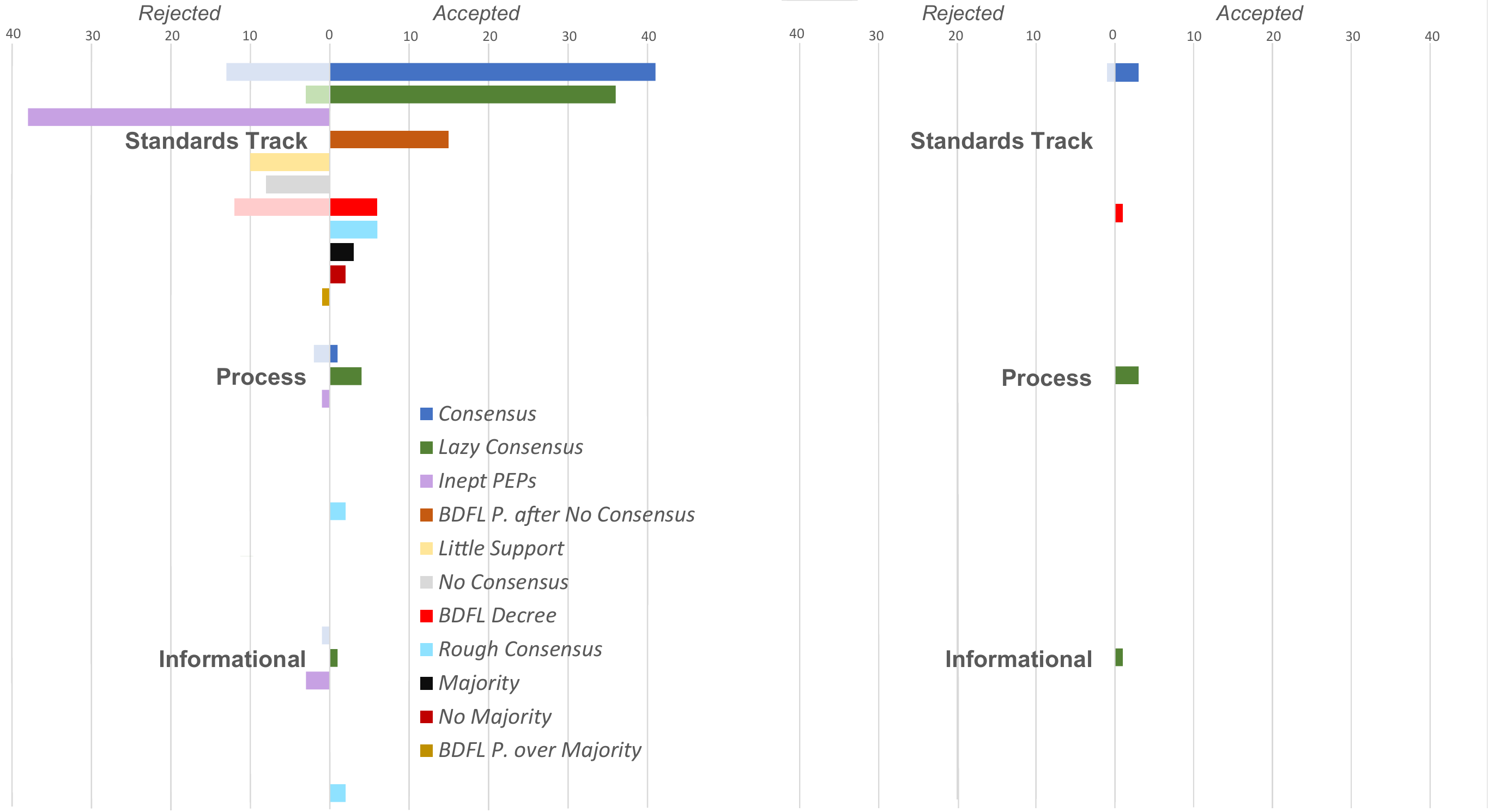}
	\end{center}
	\vspace{1pt}
	\caption{PEP decisions rationale usage by BDFL (on the left) vs.\ the BDFL Delegates (right), broken down by PEP type for \textit{accepted} and \textit{rejected} PEPs. Positive values correlate to \textit{accepted} PEPs and negative values to \emph{rejected} PEPs.}
	\label{fig:dmconceptspiechart2}
\end{figure*}

Most decisions on PEPs are based on the collective view of the community, i.e., mainly core developers. We believe this is the central aspect of DM in Python, where the project leader just ratifies (or rubber stamps) most of the decisions, and these decisions mostly originate from and lie with the views of the core members of the Python community, i.e., the meritocracy principle.

The substantial number of PEPs where the BDFL has exercised his own preference (\textit{BDFL decree}, \textit{BDFL pronouncement after no consensus}, and \textit{BDFL pronouncement after majority}) implies that in the Python community, the BDFL is free to exercise his choices when necessary. However, most of the decisions taken by the BDFL are based entirely on a collective community view,  
such as \textit{consensus, lazy consensus, rough consensus, majority, no majority, and little support}. The Python project leader allows the community to come to a collective view on proposal outcomes, using any of these rationale, and then he (or sometimes his delegates) appear to mostly concur with the collective view. 

We can see that, in general, the community view is rarely challenged by the BDFL. Overall, the combination of meritocratic and dictatorship principles in Python appears to be working well.  

It should be noted that the Python leader plays an important role in directing the project, and is needed to manoeuvre through the community disagreements. It has been noted that PEPs are accepted ``based on whether most core developers support the PEP'' \cite{pythondmcoredevs}, and wherever there were disputes between core developers and users, the BDFL normally took the core developers' view (e.g., \cite{python2004c}), but still makes independent decisions, e.g., PEP~326 \cite{python2014pep326} --- 
\textit{``Python is not a democracy.''} 
(\cite{python2004a,python2015c}). 

%


Fig.~\ref{fig:dmconceptspiechart2} depicts the distribution of rationale used by the BDFL (on the left) against those by BDFL Delegates  (on the right) across the three different PEP types (Standards Track, Informational, and Process) for \textit{accepted} and \textit{rejected} PEPs. The numbers for Process and Informational PEPs are lower because the total number of PEPs in these two PEP types is small. The detailed tallies of DM schemes and rationale grouped by PEP type are available online\textsuperscript{\ref{fn:repository}}.
The  top~5 rationale used by the BDFL for making decisions were \textit{consensus} (57), \textit{lazy consensus} (43), \textit{inept PEPs} (42), \textit{BDFL decree} (17), and \textit{BDFL pronouncement after no consensus} amongst developers (15). This implies that generally, the Python community's collective view on PEPs are preferred.

\par When comparing the distribution of rationale used by members with these roles, it is evident that the BDFL uses a wider variety of rationale (mostly in the Standards Track PEPs) compared to BDFL Delegates. Thus, we can infer that the BDFL is responsible for deciding on PEPs where there are discerning opinions in the community regarding the PEPs, and within these PEPs it requires him to use various different rationale to reach a decision. In addition, the BDFL makes most of the decisions where there is variability, i.e., he can reject PEPs, while the BDFL delegates generally decide on PEPs which are eventually accepted.

\section{Discussion}		\label{sec:discussion}

This study investigated the contributions of Python members in different roles during the decision-making process.
%
%

To achieve this, we investigated four research questions. 
Our results imply that \textbf{Python decision-making responsibility (RQ1)} has varied involvement from members in different roles. 
Unsurprisingly, the general trend is that relatively newer community members contribute most to PEP authorship, while 
the final decisions on PEPs are mainly made by senior members (core developers, PEP editors, and the BDFL). New proposals tend to come from new members who want new functionality to be added to the language. The BDFL allows these proposals from newer members, while decisions are made by (mainly) him or his trusted core developers (as his delegates). As we will discuss shortly, even where BDFL delegates make decisions, he maintains his influence by participating in discussions. 

%

We next focused on the \textbf{influence of members in different roles in PEP DM discussions (RQ2)}. 
We found that the Python project leader was more heavily involved than most members, not only in those PEPs where he made the decision, but also in those PEPs where he delegated others to decide. 
The PEP author is most involved in discussions for all PEPs and is the most prominent in terms of social roles during DM. The BDFL comes second, but his contribution is substantial (in the number of emails). In PEPs where the BDFL Delegates made the decisions, it is the delegates who are second in terms of involvement (next only to PEP authors), closely followed by the BDFL (in third position). These results show that the project leader is ever-present as a prominent contributor in various social roles during DM discussions. This implies that apart from being responsible for making decisions, the project leader also needs to be highly involved in community discussions during DM in a maturing OSS project with a mixed governance model.  

We next unearthed prominent members' influence in the social network to focus on their proximity and connectedness with other prominent members. In terms of \textbf{influence in the social network during DM (RQ3)}, our findings imply that the BDFL is closely linked with the proposal authors more than others (even the BDFL Delegate) when OSS enhancement decisions are made. This implies that he not only makes the decisions, but is  highly involved in the DM discussions. In the PEPs decided by a BDFL Delegate, one may conclude that the BDFL is still connected to the actual decision-makers during these decisions (i.e., they do not make decisions independently).


The SNA graphs suggest that the BDFL is the central and second most influential member during DM discussions when he makes decisions. Although his influence is surpassed by other members in discussions when he is not making the final PEP decision, he is still connected to the PEP author and the BDFL delegates. 
This influence might be necessary as he has previously made decisions for all other PEPs in the project. Based on his expertise, his high involvement in DM discussions on these select PEPs is crucial in order for the OSS project to be directed towards its objectives. This responsibility is recognised and is well adhered to by the BDFL, and thus is a crucial aspect in the guidance of the project. 

Finally, we used content analysis of emails to see how members from these roles influence DM. In terms of \textbf{decision-making structure (RQ4)}, the project leader keeps control of DM on controversial PEPs, while other members are responsible for DM on non-controversial PEPs. This emphasises the BDFL's role when the Python project was still maturing. While doing so, he shared with the community members how he made decisions and prepared them for this critical role by publicly communicating PEP DM rationale.

Our results imply that the BDFL had the most influence and was critical to the success of the project. However, there was still a community consensus gathering phase which should not be overlooked, as the collective community view is mostly preferred in PEP decision-making. At the same time, the BDFL was needed during disputes and he could overrule the community's collective view on occasions.


\section{Threats to validity} \label{sec:BSthreatstovalidity}
A threat to internal validity is that we may have missed some email messages that discussed PEPs, particularly those messages that did not mention a specific PEP number or PEP title and were not replies to these messages. This could impact the results for the first three RQs.
However, we do not believe this to be a major issue as PEPs are normally indicated by using their PEP number or their title. However, we do concede there may have been some messages that may not have used either of those indications. Also, in this work we have analysed only email messages. However, discussions could have been had on other platforms (e.g., IRC channels and Twitter). 

%

A threat to the external validity of the study is the generalisability of our results
to other OSSD communities. 
Since Python is a major OSS project, we believe our findings 
are representative of 
DM in projects with a mixed governance approach. However, we suggest broader investigations involving other OSS projects should be conducted to generalise the results we report here.

\section{Conclusions and Future Work}  \label{sec:conclusion}
This paper presents a longitudinal empirical study of how members from different roles impact decision-making in Python --- an OSS project with a mixed meritocracy and dictatorship governance approach. First, our study identified the roles of members making decisions. Our work showed that while relatively junior members
can propose OSS enhancements, the final decisions are made by senior members, i.e., core developers, proposal editors, and, of course, the project leader. Second, the study revealed the interaction-related activities pursued in different roles during DM (e.g., seeding posts). This analysis revealed that proposal authors were the most prominent, followed by the project leader. Third, using SNA analysis, the work showed that the project leader was the central node. In cases when someone else makes the decisions, the influence of these decision-makers were more than the project leader's. However, the project leader still contributes to the DM discussions and communicates with the decision-makers and  proposal authors. Fourth, in terms of group decision-making, 
we found that the project leader keeps the responsibility of making OSS enhancement decisions when there is a conflict within the community. To this end, the decisions made by other members (i.e., BDFL Delegates) were on proposals where there is less controversy. 

To the best of our knowledge, this study is the first to use a data-driven approach to 
examine the influence of OSSD community members who undertake different roles in DM discussions and their involvement in how OSS design decisions are made (i.e., GDM structure). 
Our findings thus enrich the knowledge surrounding the influence of OSSD members performing various roles in DM discussions and in how OSSD decisions are made. 

A promising area for future research is to undertake a focused study on analysing the sentiments of members (especially the core and influential ones) and their roles on a  proposal-by-proposal basis to uncover whose views influence decisions, and how these lead to an outburst of emotions by certain members (e.g., core members) against others, and maybe even to predict such scenarios. Another area 
could be to slice the data, based on time or milestones reached by the community to investigate how the influence of members in different roles changed during a certain period.

\bibliographystyle{ACM-Reference-Format}
\bibliography{Ref/InfluenceOfRolesInOSSDDecisionMaking_Python} 


\end{document}